\documentclass[twocolumn,prc,aps,showpacs,a4paper,floatfix,amssymb]{revtex4-1}

\usepackage{psfrag}
\usepackage{pstricks}
\usepackage{graphicx}
\graphicspath{{figures/}}
\usepackage{subfigure}
\usepackage{hyperref}
\usepackage{color}
\usepackage{amsmath}
\usepackage{soul}

\hypersetup{
  colorlinks,
  citecolor=[rgb]{0,0,1},
  linkcolor=[rgb]{0,0,1},
  urlcolor=[rgb]{0,0,1}}

\newcommand{\ud}{\mathrm{d}}

\begin{document}

\date{\today}
\title{Temperature dependence of nuclear fission time in heavy-ion fusion-fission reactions}
\author{Chris Eccles$^{1}$, Sanil Roy, Thomas H. Gray$^{1}$}
\author{Alessio Zaccone$^{1,2}$}
\email{az302@cam.ac.uk}
\affiliation{${}^1$Statistical Physics Group, CEB Department, University of Cambridge, CB2 3RA Cambridge, U.K.}
\affiliation{${}^2$Cavendish Laboratory, University of Cambridge, CB3 0HE Cambridge,
U.K.}

\begin{abstract}
Accounting for viscous damping within Fokker-Planck equations led to various improvements in the understanding and analysis of nuclear fission of heavy nuclei. 
Analytical expressions for the fission time are typically provided by Kramers' theory, which improves on the Bohr-Wheeler estimate by including the time-scale related to many-particle dissipative processes along the deformation coordinate. However, Kramers' formula breaks down for sufficiently high excitation energies where Kramers' assumption of a large barrier no longer holds. In the regime $T>1$ MeV, Kramers' theory should be replaced by a new theory based on the Ornstein-Uhlenbeck first-passage time method that is proposed here. The theory is applied to fission time data from fusion-fission experiments on $^{16}$O+$^{208}$Pb $\rightarrow$ $^{224}$Th. The proposed model provides an internally consistent one-parameter fitting of fission data with a constant nuclear friction as the fitting parameter, whereas Kramers' fitting requires a value of friction which falls out of the allowed range. The theory provides also an analytical formula that in future work can be easily implemented in numerical codes such as CASCADE or JOANNE4. 
\end{abstract}

\maketitle

\section{Introduction}
Highly excited heavy nuclei with $A>200$ resulting from a fusion process undergo fission in addition to particle decay. Experimental data from heavy-ion reactions have brought evidence for a variety of decay processes, besides fission, which include release of neutrons, charged particles, and $\gamma$-rays produced by the deformation modes (e.g. from Giant Dipole Resonance (GDR)) of the compound nucleus prior to scission~\cite{Vaz,Ajitanand,Hinde,Thoennessen}. 
The yields of pre-scission neutrons and $\gamma$-rays depend on the fission time scale: the longer it takes to reach the scission point, the larger the number of neutrons and $\gamma$-rays released overall in the process~\cite{Broglia,Woude,Hilscher,Paul}.
The measured excess of $\gamma$-radiation, with respect to estimates based on the Bohr-Wheeler width, $\Gamma_{f}^{BW}\approx\frac{T}{2\pi} \exp(-U_{f}/T)$, has called for the introduction of substantial nuclear dissipation effects which lead to longer fission time scales, which in turn can explain the amount of radiation measured~\cite{Butsch,Dioszegi1,Hofmann}. However, the nuclear viscosity, or nuclear friction parameter, which is key to the dissipative models, has become a matter of investigation in its own right, and various statistical mechanics approaches have been developed in an attempt to clarify the emergence of friction in nuclear dynamics~\cite{Hofmann,Mondal,Shlomo1,Shlomo2}.  

The most widely used expression for the fission width and for its inverse, the fission time $\tau_{f}=\hbar/\Gamma_{f}$, is the formula derived by Kramers for the rate of escape of a diffusive particle over a potential barrier~\cite{Kramers}. This treatment has been analysed and explored extensively in the context of nuclear fission by 
Weidenm\"uller and collaborators~\cite{Weidenmuller1,Weidenmuller2,Weidenmuller3}. In its basic form, the Kramers' expression for the mean fission time is given by  
\begin{equation}
\tau_{f}=\frac{\beta \mu}{2\pi \omega_0\omega_s} \exp\left(\frac{U_{f}}{T}\right)
\end{equation}
where $U_{f}$ is the fission barrier, $T$ is the nuclear temperature related to the nuclear excitation energy via $E_{x}=aT^{2}$, with $a$ the nuclear level density~\cite{Mottelson}. 
Furthermore, $\omega_{0}$ and $\omega_{s}$ represent the square root of the curvature of the fission energy landscape at the minimum and at the saddle point, respectively. 
The key parameter which encodes dissipation is the friction or dissipation coefficient $\beta$ which has dimensions $[s^{-1}]$, while $\mu$ is the inertia or reduced mass defined here as $\mu=A m_{av}$, where $m_{av}$ is the average mass of a nucleon. 
The above expression Eq. (1) was derived by Kramers using the Smoluchowski equation as a starting point, under a set of approximations for a potential landscape featuring a minimum (ground state) followed by a barrier along the reaction coordinate. The key assumptions in Kramers' derivation are that the ground state is thermalized in the potential well, and that the barrier is steep enough that a saddle-point approximation of the integrals is allowed where the potential is approximated to quadratic order both in the minimum and at the saddle~\cite{Kramers}.

The above formula is valid for the overdamped regime of high friction, whereas for moderate-to-strong friction the following formula is typically used~\cite{Weidenmuller1}, which was also derived by Kramers~\cite{Kramers}:
\begin{equation}
\tau_{f}=\frac{2\pi \omega_{0}}{\omega_s}\left(\sqrt{\frac{\eta^{2}}{4}+(2\pi\omega_{0})^2}-\frac{\eta}{2}\right)^{-1} \exp\left(\frac{U_{f}}{T}\right).
\end{equation}
This formula is derived from the 1D Fokker-Planck equation without the assumption of overdamped motion, but uses similar approximations of steep barrier and thermalization in the well. 

Focusing on the limit of overdamped dynamics, which is appropriate for heavy nuclei with $A>200$, we show that Kramers' formula breaks down dramatically at $U_{f}/T< 5$, and overestimates the fission time by up to a factor 27. We illustrate this effect on the example of a classical overdamped system, a dimer of two Brownian particles bonded via the Lennard-Jones potential with a cut-off, to present the problem in a more general context for which accurate numerical simulations are available~\cite{Abkenar}. 
Although the dimer dissociation phenomenon is different from fission of an initially spherical body into two fragments, its mathematical description in terms of diffusion dynamics as an activated escape process is almost entirely analogous. 

We then show that a mean first-passage model based on the Ornstein-Uhlenbeck (OU) method leads to a mean first-passage time formula that circumvents the limitations of the Kramers' approach, and provides accurate predictions of dissociation time-scale of the Lennard-Jones dimer in comparison with simulations down to the free diffusion limit $U_{f}/T \rightarrow 0$. 
We apply this method to nuclear fission and derive an analytical expression for the case of fission of heavy nuclei which is applicable down to vanishing barriers.
We then demonstrate the applicability of this method on the case of fusion-induced fission of $^{224}$Th for which data are available in the literature~\cite{Dioszegi2}, by estimating the potential energy landscape using the Lestone fast method~\cite{Lestone} with available input from the experimental system.  

As experimental studies of heavy-ion induced fission approach increasingly higher energies and low fission barriers due to the large angular momentum, it is clear that Kramers' theory, and also its most recent extensions, become inapplicable. 
This problem aggravates the already complicated interpretation of $\gamma$-ray and neutron spectra from fusion-fission reactions, where a factor 20 difference in estimating the fission time may obviously lead to huge errors in the estimate of particle and radiation yields.
The proposed framework provides a possible solution to this problem and it is hoped that its future refinements and implementation thereof in numerical codes will turn useful in the quantitative analysis of measured spectra. 

\section{Breakdown of Kramers theory at low barriers}
In this section we illustrate the general phenomenon of deviation from Kramers' estimate for the escape time of a Brownian system over a potential barrier.
This happens when the reduced barrier $U_{f}/T \lesssim 5$. The reason for the failure of Eq. (1) in this regime lies in the assumptions used by Kramers in his derivation, which are no longer valid. For shallow barriers one can no longer assume that the initial state is thermalized in the potential well, and, importantly, one cannot use the fact that the barrier is steep to justify the quadratic approximation of the well and of the barrier saddle point in the approximation of the integrals. 

From a different point of view, the failure of Kramers' theory becomes evident in the fact that it cannot recover the relevant limit for a vanishing barrier $U_{f}/T\rightarrow 0$. In this limit, the time scale of the process is equal to the time needed for the system to diffuse freely from the initial state (compound nucleus) to the finale state (fission fragments) along the deformation coordinate. This time is finite and equal to $L^{2} (T/\eta)^{-1}$ where $\eta=\beta \mu$. Here $L$ is the separation between the minimum of the well and the saddle point, while $T/\eta$ represents the diffusion coefficient. Kramers' formula Eq.(1) (and also Eq. (2)) fails to recover the free diffusion limit, because it predicts that $\tau_{f}\rightarrow \beta \mu/(2\pi \omega_{0}\omega_{s})$ as $U_{f}/T\rightarrow 0$.

In order to quantify this effect precisely, it is useful to consider a simple situation of two Brownian spheres interacting via the Lennard-Jones (LJ) potential. The minimum of the LJ potential represents the ground state, and the potential cut-off (defined as the separation beyond which $U=0$) plays the same role of the saddle point in the fission landscape, beyond which the particle effectively leaves the well. 
Hence the time needed for the particle to move from the minimum to the cut-off in the dimer dissociation problem is mathematically analogous to the fission time in a 1D overdamped description of the fission process as a thermal escape from the minimum up to the saddle point. 

The simulation, as detailed in Ref.~\cite{Abkenar}, is done by initializing $500-1000$ dimers in the molecular dynamics package LAMMPS and solving the Langevin equation for the dynamics. The cut-off was set to $q=3\sigma$ where $q$ is the coordinate measuring the separation between the two particles, and $\sigma$ is the particle diameter. The time needed to reach $r=3\sigma$ starting from a bound state was declared as the dissociation time for the process.
In Fig. (1) we recall the outcome of this analysis: the simulation data (symbols) start to increasingly deviate from Kramers' estimate (given by Eq. (1) above with $D=T/(\beta\mu)$ where $T$ is now the thermal energy $k_{B}T$) right below $U_{f}/T=6$. The discrepancy becomes very large when $U_{f}/T\sim 1$, below which the Kramers formula is off by a factor $>20$.

The correct behaviour can instead be predicted, in a parameter-free way, with the following method.

\begin{figure}
\includegraphics[width=1.01\columnwidth]{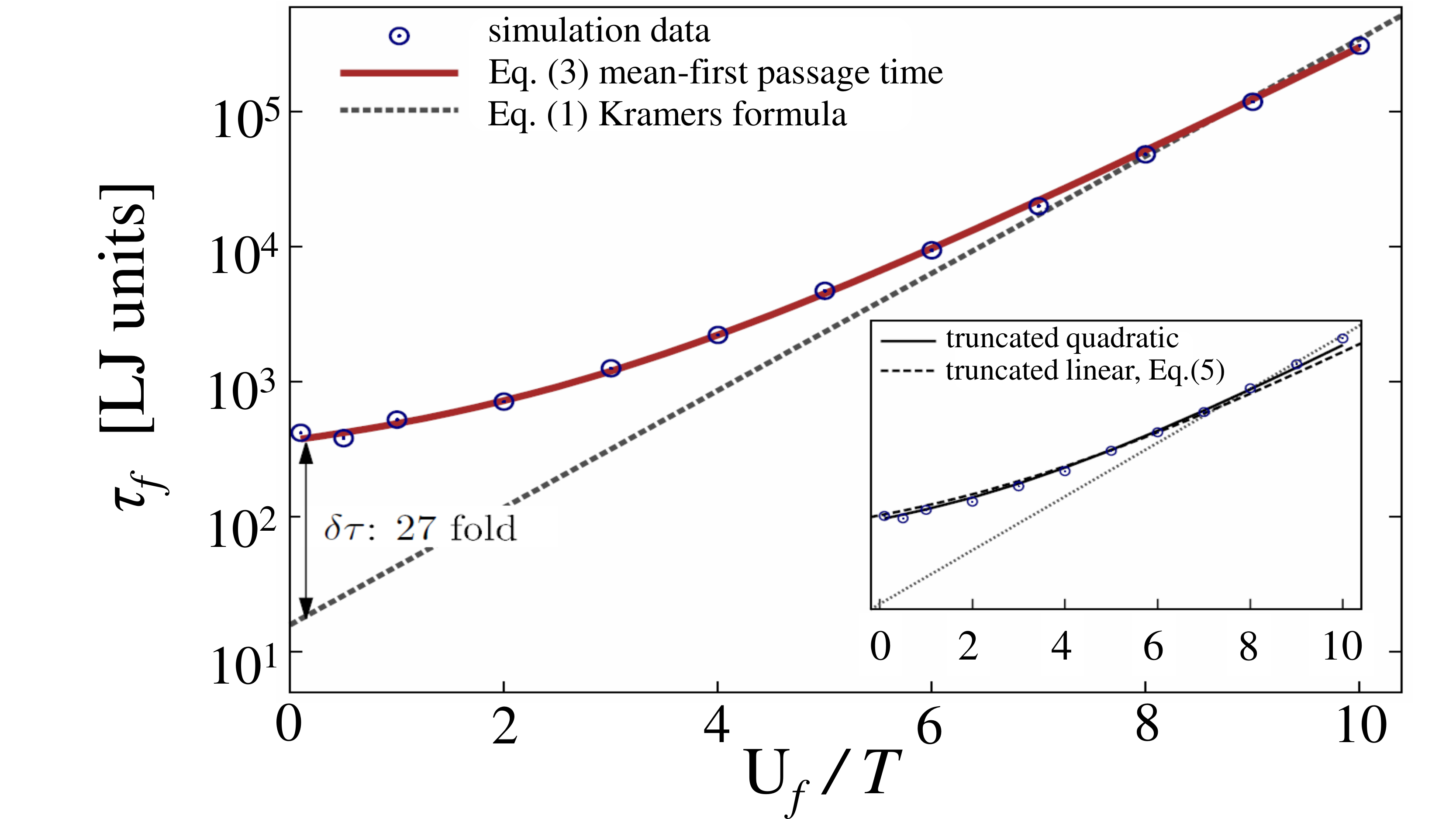}
\caption{Dissociation time vs barrier height for a dimer of spherical Lennard-Jones particles.
Kramers' Eq. (1) works well for $U_{f}/T\gtrsim7$, 
while the analytical solution, and quadratic potential model, presented here fit better across the entire range.
Errors incurred upon using the Kramers' formula can be as large as $27$-fold in the fission time. The inset shows the comparison between the same simulation data and the analytical solution Eq.(6) for the linearization approximation of Fig. 2, and also for a different approximation using a truncated quadratic well discussed in Ref.~\cite{Abkenar}. 
}
\label{Fig1}
\end{figure} 

\section{Mean-first passage time model}
Using the Ornstein-Uhlenbeck method~\cite{Uhlenbeck,Gardiner} the mean first-passage time can be determined without having to resort to Kramers' assumptions.
Instead of assuming thermalization of the bound state as in Kramers' theory, the bound state is inizialized as a delta function centred in the minimum of the well
of the potential energy landscape at $t=0$. A reflecting boundary condition is placed at $q=q_{0}$ right at the minimum of the well, while an absorbing boundary condition (sink) is placed at $q=q_{s}$, where the LJ potential is cut off. The problem is thus qualitatively the same as a fission process in the overdamped regime where the fission time is defined as the time for the system to move from the ground state (in the minimum of the well) up to the saddle. Following the derivation reported in the Appendix A, the fission time scale is evaluated according to the mean-first passage time formula 
\begin{equation}
\tau_{f} = D\int_{q_{0}}^{q_s} \ud y \exp\left(\frac{U(y)}{T}\right)\int_{q_{0}}^y \ud z \exp\left( -\frac{U(z)}{T}\right),
\end{equation}
where $D$ is the (Stokes-Einstein) diffusion coefficient of the Brownian particle (all the parameters in the comparison with the simulations are expressed in LJ units, and $T$ represents here classical thermal energy). This formula for the mean-first passage time is well known in the theory of stochastic processes and has been used before in the context of nuclear fission~\cite{Nix} although never to analyse the breakdown of Kramers formula as a function of temperature. Eq. (3) in our model is valid for a system initialized in the minimum of the well at $t=0$: this condition sets the lower limit in the outermost integral at $q_{0}$.

As discussed in detail in Ref.~\cite{Abkenar} and shown in Fig. 1, Eq. (3) produces an excellent agreement with the simulation data in a parameter-free way, down to the free diffusion limit. 

\section{Analytical formula for the fission time at arbitrary $T$}
Since the Kramers' formula has the great advantage of being in analytical form, it is important to study possible analytical versions of Eq. (3), and apply them to the case of fission. 
This can be done by approximating the potential landscape between the well minimum and the saddle point with a truncated linear or quadratic approximation. The linear approximation is schematically depicted in Fig. 2 and amounts to resetting the coordinate $q$ such that $q=0$ in the minimum, and approximating the potential between the well minimum and the absorbing boundary with a simple linear ramp,
\begin{equation}
U(q) = 
\begin{cases} 
\frac{U_{f}}{L}q & 0 < q \leq L\\
U_{f} & q > L. \\
\end{cases}
\label{eq:ramppot}
\end{equation}
Shifting the coordinate as schematically shown in Fig. 2, and placing the reflecting and absorbing boundaries at $q_0=0$ and $q=L$ respectively, produces the following expression:
\begin{equation}
\tau_{f}= \frac{L^2}{D}\frac{1}{(U_{f}/T)^2}  \left[ \exp\left(\frac{U_{f}}{T} \right) - 1\right] - \frac{L^2}{D}\frac{1}{U_{f}/T}.
\end{equation}
In the simulation of Ref.~\cite{Abkenar}, the dimers start in the potential minimum and so we start at the point lowest in potential also, which corresponds to $q=0$ in the shifted coordinate - the reflecting wall (which was $q_{0}$ before shifting the coordinate to the right to let it start from the minimum). 
As shown in the inset of Fig.1, this approximate analytical expression based on a linear approximation of the potential, provides an excellent fit of the LJ dimer dissociation data from the simulation of Ref.~\cite{Abkenar} over a broad range of $U_{f}/T$, and importantly, is able to correctly reproduce the data in the regime $U_{f}/T<5$ where Kramers' theory breaks down. 

In order to apply this formula to nuclear fission, we need to replace $D=\mu\beta/T$ and we thus obtain the following analytical expression:
\begin{eqnarray}
\tau_f=\frac{\mu\beta}{T}L^{2} \left[\frac{1}{\left(U_{f}/T\right)^2} \left(\exp\left(\frac{U_{f}}{T} \right) - 1 \right)- \frac{1}{U_{f}/T}\right].
\end{eqnarray}

\begin{figure}
\includegraphics[width=1.01\columnwidth]{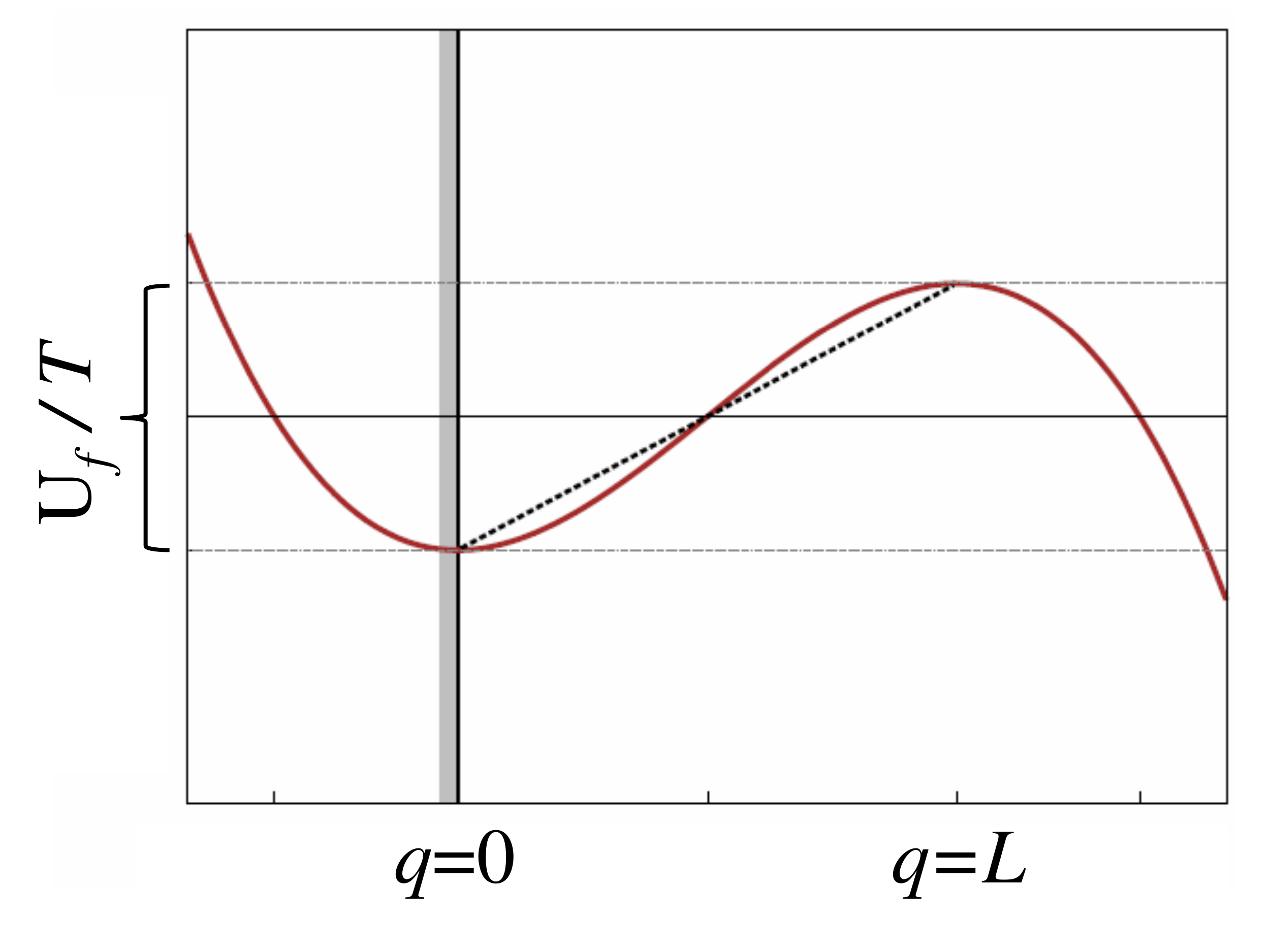}
\caption{Schematic illustration of the linearization approximation for shallow barriers. A reflecting boundary condition is placed right at the minimum of the well at $q=0$, while an absorbing boundary condition is placed at the top of the barrier (saddle) at $q=L$. Eq. (6) gives the time necessary for a system initialized at the bottom of the well at $t=0$ to reach the saddle point at a distance $L$ from the bottom of the well. 
}
\label{fig:Fig2}
\end{figure} 

An equally good fitting can be obtained with a truncated parabola instead of a linear ramp, but the resulting expression, as discussed in Ref.\cite{Abkenar}, has the disadvantage of containing hypergeometric functions. 

Equation (6) applied to nuclear fission is the main outcome of this analysis: all the parameters can be extracted from the fission energy landscape, including the length $L$ which represents the separation between the well minimum and the saddle point. Furthermore, the diffusion coefficient for the fission process is given by $D=T/(\beta \mu)$ as remarked before, where $\beta$ is the nuclear friction parameter. 

It is also important to note that for nuclear temperatures (and excitation energies) that are so large that the barrier is practically vanishing, hence for $U_{f}/T \rightarrow 0$, our model correctly yields 
\begin{equation}
\tau_f = \frac{\mu\beta}{T}L^{2} \left[\frac{1}{2} + \frac{1}{6}\left(\frac{U_{f}}{T}\right)^3 +\mathcal{O}\left\{\left(\frac{U_{f}}{T}\right)^4\right\} \right]
\end{equation}
which recovers the 1D free diffusion limit 
\begin{equation}
\tau_f=\frac{1}{2}\frac{L^{2}}{D}=\frac{1}{2} \frac{\mu\beta}{T}L^{2}.
\end{equation}

\section{Application to the $^{16}$O+$^{208}\textbf{Pb}$ $\rightarrow$ $^{224}\textbf{Th}$ system}
We now demonstrate the applicability of the above method on the example of the $^{16}$O+$^{208}$Pb $\rightarrow$ $^{224}$Th fusion-fission reaction, which is a well characterized system, and for which experimental data for the fission time are available in the literature~\cite{Dioszegi2}.

\subsection{Bound on the overdamped regime}
We shall first consider the validity of the assumption of overdamped dynamics for this system. 
Following Chandrasekhar~\cite{Chandrasekhar} and Weidenm\"uller and Jing-Shang~\cite{Weidenmuller2}, the overdamped regime, where the Fokker-Planck equation can be safely replaced with the Smoluchowski equation underlying Eq. (1) and Eqs.(3) and (6), sets in when the friction or damping coefficient is large enough that the following relations are satisfied:
\begin{align}
\beta t \gg 1 
\end{align} 
\begin{align}
\beta \gg \bigg( \frac{1}{\Delta q}\bigg) \sqrt{\frac{T}{\mu}}.
\end{align}
Equation (9) defines the time required for velocity equilibration to occur, which is a precondition to eliminate the momenta from the Fokker-Planck equation. Equation (10) states that the diffusive length scale must be small relative to the typical distance $\Delta q$ over which the potential energy landscape varies appreciably. 

Using typical values for the $^{224}$Th reaction and other heavy-ion induced fission reactions, such as $T \approx 2$MeV, $\mu \approx 224 m_{av}$ and $\Delta q \approx 1$ fm, it is found that Eq.(10) gives 
\begin{equation}
\beta \gg 4.2 \times 10^{20}s^{-1}.  
\end{equation}
This $\beta$ value provides a lower bound from which we can assume a large friction (overdamped regime) where the dynamics is governed by the Smoluchowski equation. This bound is in agreement with previous estimates~\cite{Weidenmuller1,Boilley} that typically found $\beta\geq 1.6 \times 10^{21}s^{-1}$. \\

\subsection{Estimate of the fission energy landscape} 
In order to estimate the fission energy landscape for $^{224}$Th, we employ Lestone's fast method~\cite{Lestone} which also includes the Sierk barrier correction for the liquid drop model~\cite{Sierk}. The latter accounts for the finite range of the nucleon interaction in estimating the surface energy, the Coulomb energy and the moment of inertia, by means of an empirical Yukawa-plus-exponential folding function. 
As usual in the liquid-drop model (LDM), it is assumed that the excited nucleus deformation prior to fission is dominated by quadrupolar deformation modes. These are parameterized in the spherical-harmonics expansion by $l=2$. Hence the shape is parameterized by the coordinate $q$ along which the initially spherical nucleus stretches into an oblate ellipsoid prior to necking and eventually fissioning into two spherical fragments. Hence, $q$ also represents the distance between mass centers as in the calculations reported in Ref.~\cite{Lestone}. 

In general, the energy landscape of the nucleus within the LDM has three contributions
\begin{equation}
U(q)=E_{s}(\alpha_{i})+E_{C}(\alpha_{i})+E_{r}(\alpha_{i}),
\end{equation}
$E_{s}$ is the surface energy, $E_{C}$ is the Coulomb energy, while $E_{r}$ is the rotation energy. The latter is an important contribution in heavy-ion induced reactions and directly controls the height of fission barrier $U_{f}$. The generic shape parameter $\alpha_{i}$ in our model, consistent with the tabulated data of Ref.~\cite{Lestone}, is chosen to be the normalized centre-mass distance $q/R_{0}$, where $R_{0}=r_{0}^{LDM}A^{1/3}$ is the unit of distance. For $^{224}$Th, $R_{0}=5.2$ fm.

The method makes use of several functions ($S'$, $C$, $E_S$, $I$) taken from Ref.~\cite{Lestone} and reported in Appendix B. Due to data scarcity for the present system, we use the tabulated functions for $^{208}$Pb provided in Ref.\cite{Lestone} where it is recommended, in the absence of relevant data, to approximate the potential energy landscape using these data. This estimate for the fission energy landscape gives:
\begin{widetext}
\begin{equation}
U(q, Z, A, J) = S'(q)E_s^0(Z, A) + 0.7053 C(q)\frac{Z^2}{A^{1/3}} + \frac{J(J+1)\hbar^2}{2[I_sh(q)\frac{2}{5}\mu r_0^2A^{2/3} + 4\mu a^2 ]} 
\end{equation}
\end{widetext}
where $a = 0.70$ fm and $r_0 = 1.16$ fm. Furthermore, $S'$ is a function which contains the Sierk finite-range correction for the surface energy, $C(q)$ is the Coulomb energy of a sharp-surfaced nucleus, $I_sh(q)$ is the moment of inertia determined assuming a sharp-surfaced nucleus. 

LDM energy landscapes for different values of the total angular momentum $J$ are plotted in Fig. 3. 

\begin{figure}
\includegraphics[width=1.01\columnwidth]{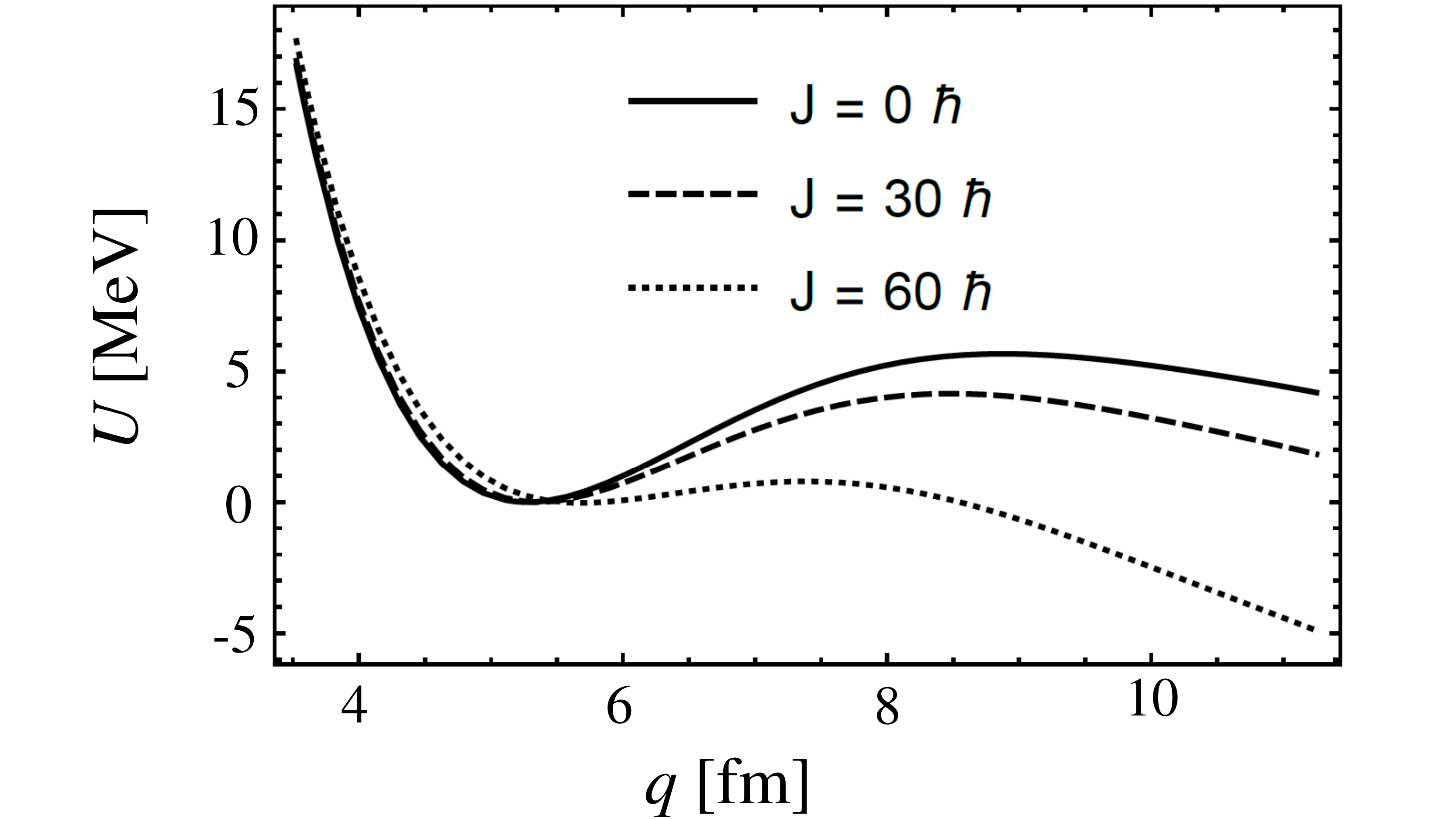}
\caption{LDM energy landscapes calculated using Eq.(11) with parameters appropriate for $^{224}$Th or when not available, with the suggested parameters for heavy nuclei reported in Ref.~\cite{Lestone}. 
}
\label{fig:Fig3}
\end{figure} 

Since the value of the angular momentum $J$ of the compound nucleus for the system $^{16}$O+$^{208}$Pb $\rightarrow$ $^{224}$Th under the conditions where fission time was extracted in Ref.~\cite{Dioszegi2} has not been reported, in the following calculations of the fission time we use $J=25 \hbar$. This value, besides complying with the usual Nordheim rules for the angular momentum of nuclei~\cite{Nordheim}, allows us to have a fission barrier equal to $U_{f}=4.58$ MeV which matches the value reported in the literature for this system~\cite{Aziz}.
 
\subsection{Comparison with experiments and discussion} 
Using Eq. (13) we can determine the fission time as a function of $T$ for $^{224}$Th under the experimental conditions of Ref.~\cite{Dioszegi2}, using both the Kramers' formula, Eq. (1), the OU mean-first passage time formula, Eq. (3), and its analytical version, Eq. (6). The latter expressions are evaluated with the boundary conditions of our model (Fig. 2). 
The comparison between the two theoretical calculations and the experimental data is shown in Fig. 4. The only adjustable parameter in the expressions used in the fittings is the friction coefficient, $\beta$. Although this is an adjustable parameter, it has to belong to the allowed range specified by Eq. (10).
The comparison in Fig. 4 thus presents calculations made using the best fitting value of $\beta$ that complies with Eq. (10). 
In particular, the best fitting using the Kramers formula Eq. (1) is obtained with $\beta=6.5 \times 10^{20} s^{-1}$. Therefore, the value of $\beta$ that would be required for Kramers' formula to fit through the data violates the condition set by Eq. (10).
Instead, the fittings using Eq. (3) and Eq. (6) are obtained with $\beta = 2.2 \times 10^{21} s^{-1}$ and $\beta = 2.5 \times 10^{21} s^{-1}$, respectively, which fulfil Eq. (10) and appear very reasonable in the context of previous studies in this regime~\cite{Weidenmuller1,Boilley}.

At low $T$ where $U_{f}/T \gg 1$, our model tends to join Kramers' formula, as expected since in that regime the Kramers assumptions become valid. At $T >2$MeV the deviation between Eq. (1) and Eq. (3) is important: Eq.(3) predicts a stronger dependence on $T$ of the fission time than the Kramers' formula. 
Although with the experimental data at hand it is not possible to further investigate this regime, previous experiments on different systems~\cite{Hinde} give evidence for fission times below $10^{-20}$ $s^{-1}$ and continuously decreasing with excitation energy even in the fast fission limit, in agreement with the expectation that the fission time must ultimately vanish in the infinite temperature limit. This is something that the Kramers' formula cannot capture since it asymptotically goes to a constant finite value in the $T\rightarrow \infty$ limit.

It is important to remark that the boundary conditions of our model for Eq. (3), and for the analytical formula Eq. (6), provide an estimate of the fission time as the time required for the nuclear deformation starting in the ground state (the minimum of the well, where we place a reflecting boundary) to reach the saddle point (where we place an absorbing boundary). This assumption is fully consistent with the definition of fission time in the experimental study of Ref.~\cite{Dioszegi2} where the fission time was measured as the time to reach the saddle point starting from the ground state, and care was put in separating this time scale from the time scale of the saddle-to-scission process. 

Finally, we should discuss the assumptions underlying the fitting in Fig. 4. An excellent fitting using Eq. (6) has been obtained with a constant value of friction $\beta$, and no need was found for including any $T$-dependence or $q$-dependence of $\beta$. 
The actual temperature dependence of the friction is still an open issue, with various models that have been proposed in the past, often with conflicting predictions. 
According to some models, a plateau in the friction at high $T$ should be reached~\cite{Hofmann}.
It is clear that with the breakdown of the Kramers' assumptions of high barrier and of thermalization in the well, and the resulting potentially large errors, assessing the $T$-dependence of friction using the Kramers' formula may easily lead to erroneous outcomes. 

We also assumed that the level density parameter $a$ remains constant during the deformation process and again this assumption did not seem to affect the quality of the fitting. In the future, the approach presented here can be further modified to include the change of the level density parameter with the deformation coordinate $q$. This point is briefly discussed in the next section.

\begin{figure}
\includegraphics[width=1.01\columnwidth]{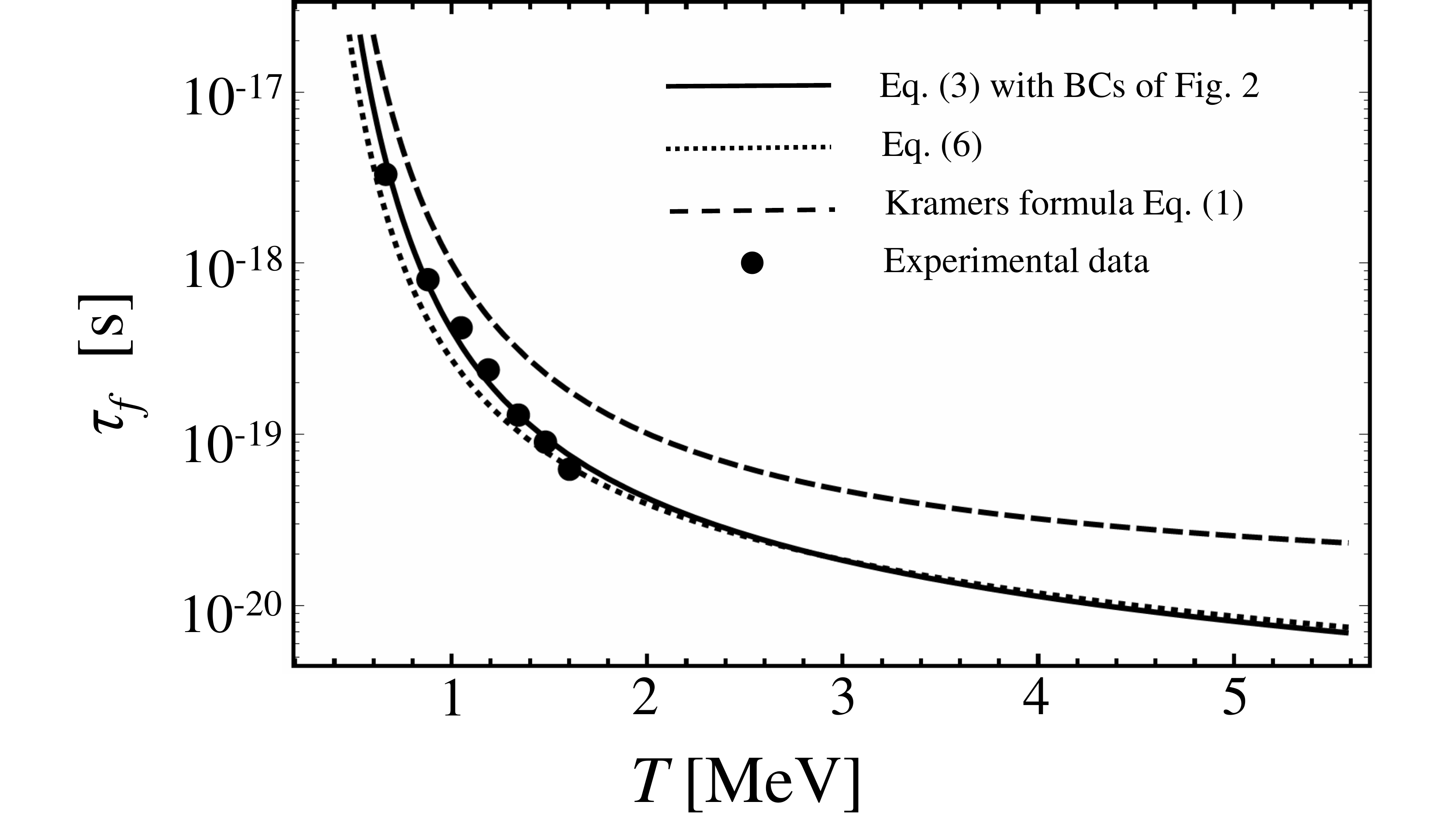}
\caption{Comparison between theoretical expressions for the fission time scale and the experimental data for the $^{224}$Th system of Ref.~\cite{Dioszegi2}. For this system the fission barrier is $U_{f}=4.58$MeV ~\cite{Aziz}. 
}
\label{fig:dudko}
\end{figure} 

\section{Accounting for $T$-dependence of energy landscape according to Refs.\cite{Lestone2008,Lestone2009}}
As was already suggested in Refs.~\cite{Frobrich,Charity}, at high excitation energy an important role in determining the shape of the energy landscape may be played by 
the deformation dependence of the level density parameter, $a(q)$. This effect can be taken into account by replacing the potential energy landscape $U(q)$ with 
an effective energy surface that contains an entropic temperature-dependent correction given by $-a(q)T^{2}$. In recent work, Lestone and McCalla~\cite{Lestone2008,Lestone2009} derived an extended Kramers model for the fission time which takes these effects fully into account. In the overdamped regime, their extended Kramers formula reads
\begin{equation}
\tau_{f}=\frac{\beta \mu}{2\pi \omega_0(T)\omega_s(T)} \exp\left(\frac{U_{f}(T)}{T}\right)
\end{equation}
where $\omega_0$, $\omega_s$ and the fission barrier $U_{f}$ are all now temperature dependent quantities since they are calculated based on the effective potential with the $-a(q)T^{2}$ correction.
This formula Eq. (14) will still suffer from the same problems due to the Kramers high barrier and thermalization assumptions, as discussed above, and may lead to  large errors in the regime $U_{f}/T<5$. 

Future work should therefore be addressed to combining our modification of Kramers theory presented above, in particular in the form of Eq. (6), which gives the correct temperature dependence for a temperature-independent energy landscape, with the Lestone and McCalla correction~\cite{Lestone2009} for the temperature-dependent landscape, to obtain the ultimate description of fission time and fission width in heavy-ion fusion-fission reactions.

\section{Summary}
Kramers' formula for the nuclear fission time scale in the overdamped regime is still widely used as an extension to the Bohr-Wheeler fission width to account for large damping in heavy-nuclei fission processes such as heavy-ion induced fusion-fission reactions~\cite{Lestone2009}. 
However, the underlying assumptions of large fission barrier and thermalization of the compound nucleus in Kramers' theory have not been properly investigated, especially in the regime of shallow fission barrier/high temperature where these assumptions break down.

Here we have shown, first on the example of a classical fission process of a Lennard-Jones dimer, that the temperature dependence of the fission time qualitatively and quantitatively deviates from the Kramers dependence $\sim \exp(U_{f}/T)$ starting already from $U_{f}/T \simeq 5$. The true fission time follows a much weaker dependence on $U_{f}/T$ in this regime, and eventually flattens out to recover the free diffusion limit as $U_{f}/T \rightarrow 0$. Kramers' formula is unable to recover the correct free-diffusion limit and gives an error of up to a factor 27.

Simulation data for the dissociation of Lennard-Jones dimers of Brownian particles are accurately reproduced by a mean-first passage time formula (based on the Ornstein-Uhlenbeck method) with boundary conditions given by a reflecting wall at the minimum of the potential energy and an absorbing boundary at the saddle point. We have shown that an analytical formula, Eq. (6), can be derived by linearizing the potential between the minimum and the saddle, which is also in excellent agreement with the data.

We applied this approach to the heavy-ion induced fission of $^{224}$Th following the $^{16}$O+$^{208}$Pb $\rightarrow$ $^{224}$Th reaction. For this well characterized system, state of the art data of the fission time are available in Ref.~\cite{Dioszegi2}. After putting a constraint on the value of the nuclear friction parameter based on the standard reduction of Fokker-Planck equation to Smoluchowski diffusion equation, we have shown that the mean first-passage time model provides an excellent fit of the data with a constant friction coefficient as the only fitting parameter. The resulting value of friction $\beta = 2.2 \times 10^{21} s^{-1}$ complies with the required bound calculated for $^{224}$Th at the conditions of the experiment. Kramers' formula is instead unable to fit the data with values of friction in the allowed range. Therefore it cannot provide an internally consistent fitting of the data. 

In conclusion, the results presented here call for a shift of paradigm and for a substantial revision of the current models of heavy nuclei fission based on Kramers' theory. This is especially important because Kramers' formula leads to both qualitatively and quantitatively erroneous estimates of the fission time and its temperature dependence.
The proposed Ornstein-Uhlenbeck mean first passage time model, for which we also provide a useful and accurate analytical expression, Eq. (6), is instead able to provide the correct temperature dependence including the limit of vanishing barrier, and appears to be accurate for the system considered here. In the future, the analytical expression Eq. (6) for the fission time proposed here should be further improved by taking into account the temperature dependence of the level density parameter in the effective potential landscape along the lines of Ref.~\cite{Lestone2008,Lestone2009}. The proposed framework will thus lead to improved expressions that can be implemented in numerical codes such as CASCADE ~\cite{Puehlhofer} and JOANNE4 ~\cite{Lestone1999}.

\begin{acknowledgements}
THG gratefully acknowledges financial support from EPSRC for his PhD studentship at University of Cambridge.
\end{acknowledgements}

\appendix
\section{Derivation of Eq. (3)}
 A large friction suggests that the Brownian, random forces acting on the system in the well are significantly larger than the forces due to the external potential $K(q)=-\partial U/\partial q$. Assuming that the potential energy landscape does not vary much over the characteristic length $\frac{1}{\beta} \sqrt{\frac{T}{\mu}} $ it can be expected that, on times larger than $1/ \beta$ we can ignore the effects of the system's initial momentum, $p$: 
\begin{align}
\rho_{FP} (q, p, t) \cong \rho_{S} (q, t) e^\frac{-p^2}{2T}.
\end{align}
Here $\sigma (q, t)$ is the density of particles in $q$ space, while $\rho$ is the total probability density. Under this assumption, a Fokker-Planck equation, which is an equation of conservation for $\rho$, reduces to the following Smoluchowski diffusion equation, which is an equation of conservation for $\rho_{S}$:
\begin{align}
\frac{\partial \rho_{S} (q,t)}{\partial t} = - \frac{\partial}{\partial q} \bigg( \frac{K(q)}{\beta \mu} \rho_{S} - \frac{T}{\beta \mu} \frac{\partial \rho_{S} (q,t)}{\partial q} \bigg).
\end{align}
Expanding the force $K(q)$ due to the potential landscape, gives:
\begin{align}
\frac{\partial \rho_{S} (q,t)}{\partial t} = \frac{\partial}{\partial q} \bigg( \frac{1}{\beta \mu} \frac{\partial U(q)}{\partial q} \rho_{S} (q,t) \bigg) + \frac{\partial}{\partial q} \bigg( \frac{T}{\beta \mu} \frac{\partial \rho_{S} (q,t)}{\partial q} \bigg).
\end{align}
The Smoluchowski Eq.(A2) is a special case of the general forward Fokker-Planck equation as given below. In our method we initialise the system from $t=0$ at the location $q_0$, 
$\rho_{S}(q,0 \rvert q_0,0) = \delta(q-q_0)$, which gives
\begin{align}
\frac{\partial \rho_{S}(q,t  \rvert q_0,0)}{\partial t} =& \frac{\partial}{\partial q} \left[-A(q) \rho_{S} (q,t  \rvert q_0,0)\right] \nonumber \\
&+ \frac{1}{2}\frac{\partial ^2}{\partial q^2}\left[B \rho_{S}(q,t  \rvert q_0,0)\right]. 
\end{align}

Assuming that $\beta$ and $\mu$ are independent of $q$, we can compare coefficients between (A4) and (A2) and write:
\begin{equation}
A(q) = - \frac{1}{\beta \mu} \frac{\partial U(q)}{\partial q}
\end{equation}
\begin{equation}
B = \frac{2T}{\beta \mu}.
\end{equation}

Equation (A4) serves as a starting point for deriving the mean first passage time formula Eq. (3) using the Ornstein-Uhlenbeck (OU) method~\cite{Uhlenbeck}. 

To calculate the probability distribution $W(q,t)$ of the particle being present in the potential energy well at time $t$ we need to integrate the probability density function across the well thus over the range $q_{0}$ to $q_{s}$:
\begin{align}
W(q_0,t) = \int_{q_{0}}^{q_{s}} \rho_{S}(q, t  \rvert q_0,0) dq.
\end{align}
If the particle is still within the well at time $t$ we know that the escape time out of the well, $\tau$, must be greater than $t$. Thus this integral also finds the probability distribution for exit times.  $W(q_0,t) = P(\tau(q_0) > t)$. The mean of this distribution yields the mean escape time from a starting location $q_0$,  as 
$\tau (q_0) $. Considering that the system is only defined from $t=0$, which is the time at which the compound nucleus is formed, we find:
\begin{align}
\tau(q_0) = \int_{0}^{\infty}W(q_0,t)dt = \int_{0}^{\infty}P(\tau(q_0) > t)dt.
\end{align}
We note that this equation is the \textit{forward} version of the Smoluchowski equation~\cite{Gardiner} because we specify the state of the system at some time and aim to discover the state of the system at some later time. The forward equation is given by:
\begin{align*}
\frac{\partial \rho_{S}(q,t  \rvert q_0,0)}{\partial t} =& \frac{\partial}{\partial q} \left[-A(q) \rho_{S} (q,t  \rvert q_0,0)\right]\\
& + \frac{1}{2}\frac{\partial ^2}{\partial q^2}\left[B \rho_{S}(q,t  \rvert q_0,0)\right].
\end{align*}

The OU method \cite{Gardiner} requires the use of the \textit{backward} Smoluchowski equation: in the backward case we know that at a future time $\tau(q_0)$ the particle leaves the well $\rho_{S}(q, \tau \rvert q_0 ,0) = 0$. This is the 'terminal condition'. Now the aim is to determine what the system's distribution was at an earlier time $t$. With the $A(q)$ and $B$ unchanged we write the backward Smoluchowski equation, or Kolmogorov equation \cite{Gardiner}:
\begin{align}
\frac{\partial \sigma(q,t \rvert q_0,0)}{\partial t} =& A(q)\frac{\partial \sigma(q,t \rvert q_0,0)}{\partial q} \nonumber \\
&+ \frac{1}{2}B\frac{\partial^2 \sigma(q,t \rvert q_0,0)}{\partial q^2}.
\end{align}
Integrating the probability density function across the well with respect to $q$ yields a differential equation in terms of $W(q_0,t)$:
\begin{align}
\frac{\partial W(q_0,t)}{\partial t} = A(q)\frac{\partial W(q_0,t)}{\partial q} + \frac{1}{2}B\frac{\partial^2 W(q_0,t)}{\partial q^2}.
\end{align}
Integrating over all time $(0, \infty$) removes the time dependence of Eq. (A10). Using Eq. (A8) to define the mean escape time we find:
\begin{align}
P(\tau(q_0) > \infty) - P(\tau(q_0) > 0) = A(q) \frac{\partial \tau (q)}{\partial q} + \frac{1}{2} B \frac{\partial ^2 \tau(q)}{\partial q^2}.
\end{align}
It is evident that the LHS of the above is equal to $-1$:
\begin{align}
-1 = A(q) \frac{\partial \tau (q_0)}{\partial q} + \frac{1}{2} B \frac{\partial ^2 \bar{\tau} (q_0)}{\partial q^2}.
\end{align}
Rearranging Eq. (A12) into the following form:
\begin{align}
\frac{\partial ^2 \tau (q_0)}{\partial q^2} + \frac{2A(q)}{B} \frac{\partial \tau (q_0)}{\partial q} = -\frac{2}{B}
\end{align}
makes it evident that the differential equation can be solved with an integrating factor. The integrating factor is:
\begin{align}
\psi (q) = \exp \bigg( \int  \frac{2A(q)}{B} dq \bigg).
\end{align}

We introduce two dummy variables $y$ and $z$, which both represent the deformation coordinate $q$. Applying the integrating factor:
\begin{align}
\frac{\partial}{\partial z} \bigg( \psi(z) \frac{\partial \tau(z)}{\partial z}   \bigg) = - \frac{2 \psi (z)}{B(z)}.
\end{align}
Integrating with respect to $z$, and recalling that the start position is at the ground state, $q_0$:
\begin{align}
\bigg[ \psi (z) \frac{\tau(z)}{\partial z} \bigg] ^{y} _{q_{0}} = -2 \int_{q_{0}}^{y} dz \frac{\psi (z)}{B(z)}.
\end{align}

The reflecting wall boundary condition of  $\partial _q \tau (q) |_{q=q_{0} } = 0$ cancels the $\psi (q_{0})$ term. Then dividing both sides by $\psi (y)$ and integrating over the range $(q_{A},q_{B})$ results in:
\begin{align}
\int_{q_{0}}^{q_{s}} dy \frac{\partial \tau(y)}{\partial y} = -2 \int_{q_{0}}^{q_{s}} \frac{dy}{\psi (y)} \int_{q_{0}}^{y} dz \frac{\psi (z)}{B(z)}.
\end{align}
Evaluating the mean time:
\begin{align}
\bar{\tau} (q_{0}) - \tau(q_{s}) =  2 \int_{q_{0}}^{q_{s}} \frac{dy}{\psi (y)} \int_{q_{0}}^{y} dz \frac{\psi (z)}{B(z)}.
\end{align}
Applying the absorbing wall condition of $ \bar{\tau} (q_{s}) = 0$, cancels out the term on the LHS, resulting in:
\begin{align}
\tau(q_{0}) =  2 \int_{q_{0}}^{q_{s}} \frac{dy}{\psi (y)} \int_{q_{0}}^{y} dz \frac{\psi (z)}{B(z)}.
\end{align}
Substituting the values for the functions $\psi, A, B$ results in the final form of Eq. (3):
\begin{equation}
\tau_{f} = \frac{\mu\beta}{T} \int_{q_{0}}^{q_s} \ud y \exp\left(\frac{U(y)}{T}\right)\int_{q_{0}}^y \ud z \exp\left( -\frac{U(z)}{T}\right).
\end{equation}

\section{Functions used in the LDM energy landscape Eq. (13)}
In order to build our estimate of the LDM energy landscape following the method by Lestone~\cite{Lestone}, the following functions that appear in Eq. (13) need to be evaluated. 
The surface energy of spherical system is given as
\begin{align*}
E_S = 17.944 \bigg[ 1- 1.783 \bigg( \frac{N-Z}{A} \bigg)^2 \bigg] A^{2/3}.
\end{align*}

The empirically adjusted finite range corrected surface energy derived by Sierk~\cite{Sierk} is given by
\begin{align*}
S'\bigg( \frac{q}{R_0} \bigg) = \frac{1 + U_f^{FRM} - \bigg[C\bigg( \frac{q_{s}}{R_0} \bigg) -1 \bigg](0.7053) Z^2 A^{-1/3}}{E_S}
\end{align*}
where $U_f^{FRM}$ is the Sierk fission barrier \cite{Sierk} while $q_{s}$ is the distance between mass centres at the saddle point. 

The Coulomb energy is given by
\begin{align*}
E_{C} = 0.7053  C(q)\frac{Z^2}{A^{1/3}} 
\end{align*}
where the function $C(q)$ is tabulated by Lestone~\cite{Lestone} for the case of $q$ being the center mass distance in quadrupolar deformation mode.

The moment of inertia for the sharp-surface spherical system is given by \cite{Davies}
\begin{align*}
I_{sh} (q) = \rho_s \int d^3 r (x^2 +y^2)
\end{align*}
where  $\rho_s$ is the mass density of the spherical system.
 
The moment of inertia corrected for the finite-range nuclear force according to Sierk~\cite{Sierk}  is given by
\begin{align*}
I \bigg( \frac{q}{R_0} \bigg) = I_{sharp}(q) \frac{2}{5} \mu R_{0}^2 + 4 \mu a^2
\end{align*}
where and $a=0.70$ fm is the range of the nuclear force~\cite{Lestone}.

\end{document}